\documentclass[preprint2]{aastex}
\usepackage{psfig,natbib}

\newcommand{\oqkev}{$1\over4$~keV}
\newcommand{\tqkev}{$3\over4$~keV}
\newcommand{\rosat}{{\it ROSAT}}

\newcommand{\rass}{{\it RASS}}

\newcommand{\einstein}{{\it Einstein}}

\newcommand{\snowden}{\mbox{$10^{-6}$ counts s$^{-1}$
arcmin$^{-2}$}}

\slugcomment{Submitted to the Astrophysical Journal}

\shortauthors{Kuntz \& Snowden}
\shorttitle{Stars and the SXRB}

\begin{document}

\title{On the Contribution of Unresolved Galactic Stars
to the Diffuse Soft X-ray Background}

\author{K. D. KUNTZ\altaffilmark{1,2}}
\affil{Department of Astronomy, University of Maryland}
\affil{College Park, MD 20742}
\email{kuntz@astro.umd.edu}
\altaffiltext{1}{Mailing address: Laboratory for High Energy Astrophysics, Code 662,
NASA/GSFC, Greenbelt, MD 20771}
\altaffiltext{2}{Current Institution: Department of Physics, University of Maryland,
Baltimore County, 1000 Hilltop Circle, Baltimore MD 21250}
\author{S. L. SNOWDEN\altaffilmark{3}}
\affil{Laboratory for High Energy Astrophysics}
\affil{Code 662, NASA/GSFC, Greenbelt, MD 20771}
\email{snowden@lheavx.gsfc.nasa.gov}
\altaffiltext{3}{Universities Space Research Association}

\begin{abstract}
Using stellar luminosity functions derived from \rosat\ data,
the contributions of Galactic stars
to the diffuse X-ray background
are calculated for \rosat\ PSPC energy bands.
The model follows that of \citet{ghmr96},
but uses \rosat\ rather than \einstein\ data
to determine the intrinsic luminosity distributions.
The model adequately predicts the numbers 
of stellar sources observed in deep \rosat\ surveys.
The contribution of unresolved stellar sources
to the \rosat\ All-Sky Survey at the Galactic poles
is 6.85, 4.76, and 4.91 $\times$ \snowden\ 
in bands R12 (\oqkev ), R45(\tqkev ), and R67(1.5 keV), respectively,
which is equivalent to 4.66, 31.3 and 26.9 $\times10^{-14}$
ergs cm$^{-2}$ s$^{-1}$ deg$^{-2}$.
\end{abstract}


\section{Introduction}

\subsection{Motivation}

Recent advances in the study of the soft X-ray background
have attempted to place limits on the amount of emission
from diffuse extragalactic sources.
\citet{co99} suggested that such a background might exist
in the \oqkev\ and \tqkev\ range as a result of gas accretion
on sub-cluster size objects forming filamentary structures
on cosmological scales,
the exact form of these structures being sensitive
to the cosmological model.
At higher energies (above $\sim1$ keV)
truly diffuse extragalactic X-ray emission is ruled out
as such gas would produce unacceptably large perturbations
in the microwave background \citep{wmfk94}.
Such considerations do not hold at \oqkev ,
but the multiplicity of backgrounds
(at least two diffuse Galactic backgrounds,
as well as unresolved Galactic point sources,
and unresolved extragalactic sources)
make the detection of a diffuse extragalactic background difficult.
\rosat\ observations have been used to attack the problem
on several different fronts.

The extragalactic background due to unresolved point sources
can be characterized as a power-law at E$>$1 keV,
and a variety of results, including recent Chandra observations
\citep{mcba2000}
suggests that this power-law extends,
without a break,
to lower energies.
\citet{sfks2000} and \citet{ks2000},
following the work of \citet{sealhb98},
have studied the emission that is not
due to unresolved extragalactic sources.
\citet{sealhb98} separated the \oqkev\ emission
originating from beyond the Galactic absorption
from that due to the Local Hot Bubble,
and this work was extended by \citet{ks2000}
to the remainder of the \rosat\ band.
Although \citet{ks2000} characterized the emission
originating beyond the Galactic absorption
in the \tqkev\ band,
they were not able to effect a separation
of the emission from beyond the Galactic absorption
into its various components: 
the Galactic halo, the diffuse extragalactic emission,
and unresolved Galactic point sources.
Thus, although there is almost no doubt 
that there is a Galactic halo,
the upper limit to the diffuse extragalactic emission
in the \tqkev\ band remains
the total emission originating beyond the Galactic absorption.
Included in the measure of the emission
originating beyond the Galactic absorption
was the bulk of the contribution due to unresolved Galactic point sources.
It is thus necessary to calculate that contribution
to further tighten the limits on the diffuse extragalactic background.

From the model of \citet{co99},
the diffuse extragalactic background will not be smooth,
and the clumpiness is a function of the cosmological model.
\citet{kuntz2000}
sought to place limits on the spatial fluctuations
that might be due to the diffuse extragalactic background,
thus it was necessary to determine the degree of fluctuation
due to unresolved Galactic point sources.

It is thus particularly interesting
for the study of the diffuse extragalactic background
to study the luminosity function of Galactic point sources,
which, at high Galactic latitudes, are stars.

\subsection{Previous Work}

A stellar log N-log S relation 
in the form of a cumulative luminosity function
(CLF, the number of sources, N,
with fluxes greater than S)
has been calculated by \citet{ghmr96}
from a mixture of {\it Einstein} IPC and \rosat\ PSPC data,
using the \citet{brc87} model of the spatial distribution of stars.
The \citet{ghmr96} study concentrated on comparing the model
distribution of stellar sources with observed count rates
as a function of position,
the complement of the information required for studies
of the diffuse X-ray background.

\begin{figure*}[t!]
\centerline{\psfig{figure=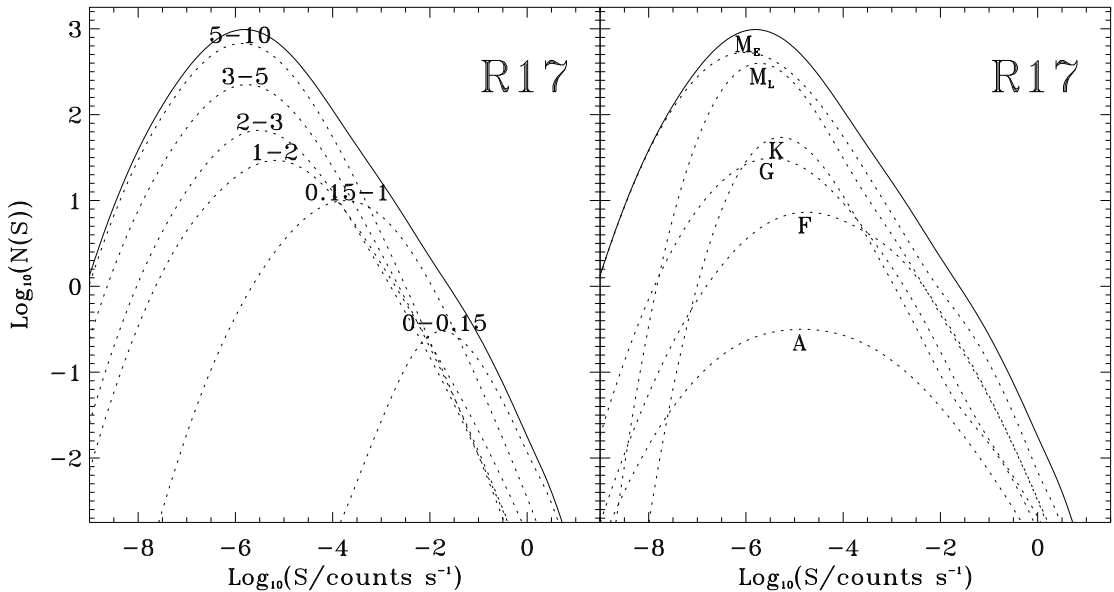,width=16.0cm}}
\caption{The luminosity function, $N(S)d\log{S}$, for band R17
shown with the fractional contributions of the
different age classes ({\it left})
and spectral types ({\it right}).}
\label{fig:lnls17}
\end{figure*}

There are a number of difficulties in applying the published
\citet{ghmr96} CLF to the problem of unresolved
stellar sources.
First, the \citet{ghmr96} study published the CLF
for the entire \rosat\ PSPC band-pass.
Since the angular structure of the diffuse X-ray background
is strongly energy dependent,
studies of the diffuse X-ray background are generally
confined to only some portion of that band-pass.
Given the strong differences in the absorption by the ISM
from one end of the band-pass to the other,
no simple scaling can be applied to the published relations
to obtain the relation for any sub-band.
Second, as can be seen in \citet{ghmr96},
and again in our Figure~\ref{fig:lnls17},
although older stars have lower X-ray luminosities,
they form such a large fraction of the total stellar population
that they dominate the luminosity function,
especially at lower flux levels.
However, this is the population that is the least
well characterized in the \citet{ghmr96} study;
they were forced to take data from {\it Einstein} IPC studies
of field stars.
Far superior \rosat\ studies have since become available
\citep{sfg95,sfg97}

The following model calculation uses the best current data
to derive the luminosity function at high Galactic latitude,
which will be of greatest interest to those studying
the diffuse extragalactic background.
Rather than the CLF, we calculate the more useful form,
\begin{equation}
N(S)d\log{S},
\end{equation}
which allows the calculation of 
\begin{equation}
\int^{S^{\prime}}_0 SN(S)d\log{S},
\end{equation}
the flux in unresolved sources.
The model is, in principle, capable of calculating
the stellar contribution at low Galactic latitudes,
but some poorly determined input parameters,
the intrinsic luminosity probability distribution for young stars,
greatly increase the uncertainties near the Galactic plane.
The irregularity of the spatial distribution of the ISM
also introduces some uncertainty,
more in softer bands than in harder,
and more at lower Galactic latitudes.

\section{Data}

In order to calculate the contribution of stellar sources,
the stellar populations were binned by spectral type ${\cal S}_T$
and age ${\cal A}$.
We used the spectral-type divisions
of \citet{sfg95} and \citet{sfg97}
and the age divisions of \citet{ghmr96},
all of which are listed in Table~\ref{tab:stardat}.
The division by age is as follows:
the 0-0.15 Gyr class is represented by
the Pleiades (171 \rosat\ detected members, $\sim0.1$ Gyr, 127 pc),
the 0.15-1.0 Gyr class is represented by
the Hyades (191 detected members, $\sim0.6$ Gyr, 45 pc),
and the 1.0-10.0 Gyr class is represented by the field population.

The data are insufficient to determine
the X-ray luminosity functions for subsamples
of the 1-10 Gyr field population,
which, especially for the later type stars,
will be nearly uniformly distributed through the age range.
However, stars in the 1-2 Gyr age range
will have a very different {\it z}-height distribution
than stars in the 5-10 Gyr age range.
Thus, for calculation, the stars in the 1-10 Gyr age class
are subdivided into the age ranges used by \citet{brc87}
(see Table~\ref{tab:starnum})
although the same X-ray luminosity function is used
for all the age sub-divisions in the 1-10 Gyr age class.

A number of other well-studied clusters could be included
with the Hyades for a sample of stars in the 0.15-1.0 Gyr group;
Coma Berenices (42 detected members, $\sim0.5$ Gyr, 80 pc) 
and Praesepe (56 detected members, $\sim0.6$ Gyr, 160 pc).
X-ray emission from Coma Berenices members
is similar to that from Hyades members \citep{rsp96},
while that from Praesepe members 
is lower than that from (coeval) Hyades members \citep{rs95}.
Given that the Hyades would seem to be representative
of its age class,
and that there is a relatively larger sample 
available from the Hyades cluster
than from other clusters,
the other clusters have not been included in this age class.
However, it should be borne in mind that the Hyades luminosities
may need to be adjusted downwards to truly represent its age class.

Although X-ray emission has been detected
from nearly every spectral type,
the dominant contribution by stellar sources
is from low-mass main sequence dwarfs,
due both to intrinsic luminosity
and the number of sources.

Stars of spectral type A have been excluded from the model.
The intrinsic X-ray luminosity probability distribution
is poorly determined for the younger age classes,
and is currently undetermined for the 1-10 Gyr age class,
which would dominate the emission
from the A spectral type.
The intrinsic X-ray luminosities, where measurable,
are about an order of magnitude below those of F stars,
and the mid-plane number density is lower than F stars,
by a factor of 0.005 for the 1-10 Gyr age class.
Since the younger age classes will have low {\it z}-heights,
the error introduced will be minimal for high Galactic latitudes.
\citet{ghmr96} estimate the maximum contribution from A stars
to be $\sim2$\%.

White dwarfs have also been excluded from the calculation
as their contributions are typically two orders of magnitude
below that of main-sequence stars in the \oqkev\ band,
and significantly lower in the \tqkev\ and 1.5 keV bands.

RS CVn systems have also been excluded from the calculation.
Although they have high luminosities,
the median intrinsic luminosity, $\log L_x\sim30.3$
(here, as in the remainder of this paper,
$L_x$ is in ergs s$^{-1}$ in the 0.1-2.4 keV band) \citep{dlfs93},
their number density is quite low,
probably less than $1\times10^{-4}$ pc$^{-3}$ \citep{os92}
as is their scale height ($\sim325$ pc).
At high Galactic latitudes,
any RS CVn system is likely to be close enough
to be detected and not contribute to the ``diffuse'' background.
Giants and super giants have been excluded from the calculation
as their proper calculation would require extensive analysis
beyond the scope of this paper.
Nearly all giants earlier than K3 are X-ray sources with $\log{L_X}\sim27.3$,
similar to solar type stars \citep{rhs98,hssz98}
but the source population is much smaller,
and more closely restricted to the plane.

We have also excluded Population II stars from the calculation.
If a comparison between disk RS CVn systems
and Population II binaries is any indication
of the relative emission from single
Population I and Population II stars,
then the Population II stars should have emission
that is more than an order of magnitude below
that of Population I stars \citep{ofp97}.

\begin{deluxetable}{lccccccccccc}
\rotate
\tablecolumns{12}
\tabletypesize{\footnotesize}
\tablecaption{Stellar Data
\label{tab:stardat}}
\tablewidth{0pt}
\tablehead{
\colhead{Type} &
\colhead{B-V} &
\colhead{M$_V$} &
\multicolumn{3}{c}{0-0.15 Gyr} &
\multicolumn{3}{c}{0.15-1 Gyr} &
\multicolumn{3}{c}{1-10 Gyr} \\
\colhead{ } &
\colhead{ } &
\colhead{ } &
\colhead{No.\tablenotemark{a}} &
\colhead{$\mu$\tablenotemark{b}} &
\colhead{$\sigma$\tablenotemark{b}} &
\colhead{No.\tablenotemark{a}} &
\colhead{$\mu$\tablenotemark{b}} &
\colhead{$\sigma$\tablenotemark{b}} &
\colhead{No.\tablenotemark{a}} &
\colhead{$\mu$\tablenotemark{b}$~$\tablenotemark{c}} &
\colhead{$\sigma$\tablenotemark{b}$~$\tablenotemark{c}} }
\startdata
A\tablenotemark{d}        & -0.01-0.29   & \nodata
                        &  16/39 & 28.51$\pm$0.26 & 0.99$\pm$0.16 
                        &  10/35 & 27.84$\pm$0.39 & 0.93$\pm$0.29 
                        &  16/62\tablenotemark{e} &    \nodata     &   \nodata     \nl
F        & 0.3-0.6      & \nodata
                        &  30/33 & 29.39$\pm$0.07 & 0.32$\pm$0.06 
                        &  57/60 & 28.93$\pm$0.05 & 0.28$\pm$0.03 
                        &  28/29 & 28.00$\pm$0.14 & 0.96$\pm$0.14 \nl
G        & 0.61-0.8     & \nodata
                        &  19/26 & 29.39$\pm$0.20 & 0.67$\pm$0.18 
                        &  27/33 & 28.64$\pm$0.12 & 0.39$\pm$0.05 
                        &  29/38 & 27.36$\pm$0.17 & 0.82$\pm$0.13 \nl
K        & $>0.8$       & $<8$
                        &  40/44 & 29.46$\pm$0.07 & 0.42$\pm$0.04 
                        &  51/86 & 28.29$\pm$0.06 & 0.45$\pm$0.08 
                        &  17/17 & 27.62$\pm$0.17 & 0.48$\pm$0.12 \nl
M(early) & $>0.8$       & 8-15
                        &  44/55 & 29.19$\pm$0.05 & 0.32$\pm$0.04 
                        & 56/174 & 27.89$\pm$0.14 & 0.72$\pm$0.12 
                        &  66/78 & 26.86$\pm$0.10 & 0.77$\pm$0.10 \nl
M(late)  & $>0.8$       & $>15$
                        &  0     & 29.12\tablenotemark{f} & 0.49\tablenotemark{f} 
                        &  0     & 28.20\tablenotemark{f} & 0.49\tablenotemark{f} 
                        &  16/19 & 27.22$\pm$0.22 & 0.49$\pm$0.24 \nl
\hline
\multicolumn{2}{l}{Spatial Distribution}
            &         & \multicolumn{3}{l}{Bienaym\'{e} et al. Eq. A1}
                              & \multicolumn{3}{l}{Bienaym\'{e} et al. Eq. A2}
                              & \multicolumn{3}{l}{Bienaym\'{e} et al. Eq. A2}  \nl
\hline
\multicolumn{2}{l}{Stellar Sample}
            &         & \multicolumn{3}{l}{Pleiades}
                              & \multicolumn{3}{l}{Hyades}
                              & \multicolumn{3}{l}{Field Stars} \nl
            &         & & & & & & & & \multicolumn{3}{l}{d$<$13 pc for F \& G stars} \nl
            &         & & & & & & & & \multicolumn{3}{l}{d$<$7 pc for K \& M stars}\nl
\hline
Data\tablenotemark{g}     &
            &         & \multicolumn{3}{l}{Stauffer et al. (1994), PSPC}
                              & \multicolumn{3}{l}{Stern et al. (1995), RASS}
                              & \multicolumn{3}{l}{Schmitt et al. (1995), PSPC} \nl
            &         &         & \multicolumn{3}{l}{ }
                              & \multicolumn{3}{l}{Pye et al. (1994), PSPC}
                              & \multicolumn{3}{l}{Schmitt (1997), PSPC} \nl
\enddata
\tablenotetext{a}{The number of detected stars in the sample divided by
the total number of stars in the sample.}
\tablenotetext{b}{The quantities $\mu$ and $\sigma$ are in $\log_{10}L_x$
where $L_x$ is in units of ergs s$^{-1}$ in the 0.1-2.4 keV band.
The uncertainties were determined as described in the text.}
\tablenotetext{c}{Since Schmitt (1997) and Schmitt et al. (1995) convert from
PSPC counts to luminosity using
(5.30HR + 8.31)$\times10^{-12}$ erg cm$^{-2}$ s$^{-1}$ [count s$^{-1}$]$^{-1}$
where HR is a hardness ratio,
and since HR is luminosity dependent,
the $\sigma$ derived from Schmitt (1997) and Schmitt et al. (1995) data
are likely to be systematically larger than those derived from other data sets
where the PSPC count to luminosity conversion was made with a standard
$6\times10^{-12}$ erg cm$^{-2}$ s$^{-1}$ [count s$^{-1}$]$^{-1}$.
This difference in conversion may also introduce
a very small offset between these luminosities
and those of the other age classes.
Our fit values were similar to those of \citet{sfg95} and \citet{sfg97},
and we have retained their values.}
\tablenotetext{d}{Spectral-type A is not used in the model calculations.
Note the large uncertainties in the parameters of the intrinsic luminosity distribution.
The parameters could not be calculated for the field sample
as the detected fraction for a sample of main-sequence A stars within 30 pc
is $\sim1/3$, and upper limits for the non-detections have not been published.}
\tablenotetext{e}{Number in the \rass\ Point Source Catalogue \citep{hssv99}.
In the $r<13$ pc sample of \citet{sfg97} there are four A stars,
none of which were detected.}
\tablenotetext{f}{Extrapolated from the 1.0-10.0 Gyr class
due to lack of data. See text.}
\tablenotetext{g}{``PSPC'' refers to data obtained from pointed PSPC observations,
``RASS'' refers to data extracted from the \rosat\ All-Sky Survey.}
\end{deluxetable}

\begin{deluxetable}{lcccccc}
\tablecolumns{7}
\tabletypesize{\footnotesize}
\tablecaption{Stellar Data
\label{tab:starnum}}
\tablewidth{0pt}
\tablehead{
\colhead{Type} &
\multicolumn{6}{c}{Age Group (Gyr)} \\
\colhead{ } &
\colhead{0-0.15} &
\colhead{0.15-1} &
\colhead{1-2} &
\colhead{2-3} &
\colhead{3-5} &
\colhead{5-10} }
\startdata
A        & $2.16\times10^{-4}$ & $3.93\times10^{-4}$ & $1.01\times10^{-4}$ & $2.55\times10^{-7}$ & 0 & 0\nl
F        & $2.93\times10^{-4}$ & $6.14\times10^{-4}$ & $4.40\times10^{-4}$ & $2.59\times10^{-4}$ & $2.29\times10^{-4}$ & $6.99\times10^{-5}$ \nl
G        & $2.78\times10^{-4}$ & $5.82\times10^{-4}$ & $4.17\times10^{-4}$ & $2.88\times10^{-4}$ & $4.39\times10^{-4}$ & $9.15\times10^{-4}$ \nl
K        & $3.31\times10^{-4}$ & $6.93\times10^{-4}$ & $4.96\times10^{-4}$ & $3.42\times10^{-4}$ & $5.22\times10^{-4}$ & $1.18\times10^{-3}$ \nl
M(early) & $4.28\times10^{-3}$ & $8.96\times10^{-3}$ & $6.42\times10^{-3}$ & $4.43\times10^{-3}$ & $6.75\times10^{-3}$ & $1.53\times10^{-2}$ \nl
M(late)  & $2.45\times10^{-3}$ & $5.12\times10^{-3}$ & $3.67\times10^{-3}$ & $2.53\times10^{-3}$ & $3.86\times10^{-3}$ & $8.73\times10^{-3}$ \nl
\hline
c\tablenotemark{a}& 0.0140 & 0.0279 & 0.0457 & 0.0662 & 0.0867 & 0.0958 \nl
\enddata
\tablenotetext{a}{Parameter describing the ellipticity of the stellar distribution.
See equations A1 and A2 of \citet{brc87}.}
\end{deluxetable}

\subsection{Model Inputs}

For each spectral-type/age-class bin,
the contribution to the luminosity function
is determined by four parameters:
$\mu$ and $\sigma$,
the mean and dispersion of
the intrinsic luminosity probability function,
$c$, which describes the $z$ distribution of the stars,
and $n_0$, the mid-plane ($z=0$) density of stars.

Following \citet{sfg95} and \citet{sfg97},
we have assumed that the intrinsic luminosity function
for a spectral-type/age-class bin is given 
by the log-normal distribution,
\begin{equation}
P(\log{L_x}) = \frac{1}{\sigma \sqrt{2\pi}}
\exp{-\frac{1}{2}\left[\frac{\log{L_x}-\mu}{\sigma}\right]^2}.
\end{equation}
The parameters 
$\mu(S_T,{\cal A})$ and $\sigma(S_T,{\cal A})$
were derived from the \rosat\ data sets 
listed in Table~\ref{tab:stardat}.

Since these data sets contain points with upper limits,
the parameters were determined by fitting
the Kaplan-Meier estimator \citep{schmitt85}
of the cumulative distribution function,
where the uncertainty in the Kaplan-Meier estimator
was taken from \citet{fn85}.
The uncertainties were derived
using a bootstrap method \citep{bf96}.
We corrected for binaries
in the same manner as \citet{sfg95}:
we have divided the flux equally between
members of binary systems.
\citet{ghmr96} showed that correction for binaries
made little difference to the luminosity function.

Since there were no late M stars detected
in the surveys of the Hyades or Pleiades,
we have extrapolated the $\mu$ for the
the 0-0.15 Gyr and 0.15-1.0 Gyr age classes 
from the $\mu$ for the 1-10 Gyr age class
using the general trend observed in the other spectral classes;
a $\Delta\log{L_X}=0.92$ decrease between 
the 0-0.15 Gyr and 0.15-1.0 Gyr classes,
and a $\Delta\log{L_X}=0.98$ decrease between 
the 0.15-1.0 Gyr and 1.0-10.0 Gyr classes.
For the late M stars the same value of $\sigma$ 
was used for each age class,
that of the oldest age class;
this tends to increase the number of sources
with high fluxes, but not significantly.

To convert from the listed X-ray luminosities to count-rates,
we used the inverse of the conversion used
by the original authors.
\citet{ssk95} used the conversion
$6.0\times10^{-12}$ ergs cm$^{-2}$ s$^{-1}$ 
[counts s$^{-1}$]$^{-1}$.
\citet{scgph94} used a conversion of
$\sim8.4\times10^{-12}$ ergs cm$^{-2}$ s$^{-1}$
[counts s$^{-1}$]$^{-1}$
(for $N_H = 0$), characteristic of $\log T=7.06$.
Due to the shape of the conversion curve,
this is also the conversion factor for $\log T=6.88$.
\citet{sfg95} used the conversion factor\footnote{
This relation was derived for a Raymond \& Smith model
two component corona fitted to a small number of high S/N spectra.
See \citet{fmmw95} for a complete description.}
$(5.30HR + 8.31)\times10^{-12}$ ergs cm$^{-2}$ s$^{-1}$ 
[counts s$^{-1}$]$^{-1}$,
where $HR$ is defined by (R47-R82)/(R47+R82)
and R$XX$ is the count rate in band $XX$
and the bands are defined in \citet{rass2}
and again in Table~\ref{tab:bands}.

\begin{deluxetable}{lccl}
\singlespace
\tablecolumns{4}
\tabletypesize{\footnotesize}
\tablecaption{Broad Band Definitions
\label{tab:bands}}
\tablewidth{0pt}
\tablehead{
\colhead{Band} &
\colhead{PI Channel} &
\colhead{Energy\tablenotemark{a}} &
\colhead{ } \\
\colhead{ } &
\colhead{ } &
\colhead{(keV)} &
\colhead{ } }
\startdata
$\begin{array}{l} \mathrm{R1}\\ \mathrm{R1L=R8} \\ \mathrm{R2} \end{array}$ &
$\mathrm{\begin{array}{c} 8-19\\11-19\\20-41 \end{array}}$ &
$\mathrm{\begin{array}{c} 0.11-0.284\\0.11-0.284\\0.14-0.284 \end{array}}$ &
$\left\} \begin{array}{l} \mathrm{R12}\\ \mathrm{or}\\ \mathrm{R1L2} \end{array} \right. =\frac{1}{4}~\mathrm{keV}$ \\

$\begin{array}{l} \mathrm{R3} \end{array}$ & $42-51$ & $0.20-0.83$ & \\

$\begin{array}{l} \mathrm{R4}\\ \mathrm{R5} \end{array}$ &
$\mathrm{\begin{array}{c} 52-69\\70-90 \end{array}}$ &
$\mathrm{\begin{array}{c} 0.44-1.01\\0.56-1.21 \end{array}}$ &
$\left\} \begin{array}{l} ~\\~ \end{array}\right. \mathrm{R45}=\frac{3}{4}~\mathrm{keV}$ \\

$\begin{array}{l} \mathrm{R6}\\ \mathrm{R7} \end{array}$ &
$\mathrm{\begin{array}{c} 91-131\\132-201 \end{array}}$ &
$\mathrm{\begin{array}{c} 0.73-1.56\\1.05-2.04 \end{array}}$ &
$\left\} \begin{array}{l} ~\\~ \end{array}\right. \mathrm{R67}=1.5~\mathrm{keV}$ \\
\enddata
\tablecomments{The notation band RXY
denotes the sum of all of the bands
between (and including) RX and RY.
In the \rosat\ source catalogues,
HR1=(R47-R12)/(R47+R12) and
HR2=(R67-R45)/(R67+R45).}
\tablenotetext{a}{10\% of peak response.}
\end{deluxetable}

From the band R17 count rates
it is neccessary to derive count rates in narrower bands
for this study to be of use.
To divide the band R17 into the R12 and R47 bands,
one can use the $\log L_x$-$HR$ developed by \citet{schmittcomm}.
Schmitt suggested that
for each spectral type,
$HR$ is a linear function of the log of the luminosity,
\begin{equation}
\log L_x = a HR + b;
\end{equation}
the slopes of these relations 
seem to be the same for all spectral types,
but the intersections are different.
The parameters of these relations were determined 
for the field stars from the data of 
\citet{sfg95} and \citet{sfg97} by \citet{kuntz2000}.

The applicability of these relations to younger stars is not clear.
Extracting the Hyades members identified by \citet{ssk95}
from the {\it ROSAT} Bright and Faint Source Catalogues 
\citep{rbsc99,rfsc2000},
we found that the field star relations fit the Hyades stars well
for spectral types F and G. 
For the M stars, the slopes of the relations appeared consistent,
but the field star offsets were too large (i.e., too hard)
for the Hyades stars.
Since there is a substantial dispersion around the relation
for both the field stars and the Hyades,
and there may be significant biases in the selection
of sources for the Bright and Faint Source Catalogues,
we have used the field star relations for the younger stars as well.
There are not enough Pleiades stars in the Faint Source Catalogue
to attempt a similar analysis.

To further subdivide the flux,
we investigated the relation between $HR$ and the spectral shape.
From the {\it HEASARC} archive of {\it ROSAT} PSPC exposures
we extracted the spectra of all single field dwarfs of spectral classes
A through M that had more than $\sim 400$ counts,
about 16 stars.
To these spectra we used XSPEC \citep{xspec}
to fit two \citet{rs77} components
with temperatures $T_S$ and $T_H$ and normalizations $N_S$ and $N_H$.
We found that the temperature of the softer component, $T_S$,
did not vary with $HR$ ($\log T_S\sim6.23$),
but that for stars in an individual spectral type,
the temperature of the harder component, $T_H$,
varied linearly with $HR$\footnote{
We found that the F and G stars followed one $T_H$-$HR$
relation while the M stars followed another.
There were only two K stars in the sample,
one of which fell on the relation for F and G stars,
the other of which fell between the FG and M relations.
Of the M stars, half were dM and half were dMe,
so the difference between the FG and M relations
is not due to selecting the brighter M stars.}.
Given the $T_H$ for a given $HR$,
we then solved for the $N_S/N_H$ ratio that would produce
the correct value of $HR$.
This relation between $T_H$ and $HR$ restricts the range
of allowable $HR$ to $>-0.86$ and $<0.37$ (for M stars)
and $<0.22$ (for F and G stars).

We checked this $T_H$, $N_S/N_H$, $HR$ relation
against the Hyades stars extracted from
the Bright and Faint Source Catalogues
using (R47-R82)/(R47+R82) {\it vs.} (R67-R45)/(R67+R45) graphs,
and found that within the large uncertainties
of the data, the relations derived from field stars
also described the Hyades stars.

Thus, for a given X-ray luminosity,
we used the Schmitt relation between the X-ray luminosity
and the hardness ratio to determine $HR$,
and then use these relations between $T_H$, $N_S/N_H$, and $HR$
to derive a spectral energy distribution that can then be used
to divide the total number of counts into the counts for each 
of the {\it ROSAT} bands.
Since the $\log L_X$-$HR$ relation can produce values of $HR$
ouside the range allowed by the $T_H$, $N_S/N_H$, $HR$ relation,
values of $HR$ outside this range were replaced by the closest extreme.

The spatial distribution of each spectral-type/age-class
was assumed to be given by equations A1 and A2 of \citet{brc87},
where $c({\cal A})$, 
the axial ratio for ellipsoids describing the disk populations,
were taken from the ``best fit'' column of their Table 1
and are listed in our Table~\ref{tab:starnum}.

To maintain some consistency between the initial mass function (IMF), 
the stellar formation rate (SFR),
and the $\sigma_z$-Age relation, 
the mid-plane density of each spectral-type/age-class bin,
$n_0(S_T,{\cal A})$,
was derived in a manner similar to that used by \citet{ghmr96}.
The disk was assumed to have an age of 10 Gyr,
a constant SFR, and an invariant IMF.
The IMF was taken from \citet{hrc97a}:
\begin{equation}
dN = m^{-(1+x)}dm
\end{equation}
where
\begin{eqnarray}
x=0.7~\mbox{for}&~m < m_{\sun}\nonumber \\
x=1.5~\mbox{for}&~1m_{\sun} \le m < 3m_{\sun} \\
x=2.0~\mbox{for}&~m \ge 3m_{\sun} \nonumber.
\end{eqnarray}
The total number of stars was normalized by comparing
the number produced by the above formula with $r<7$ pc
in the $8< M_V <15$ range
(i.e., the K and early M stars)
with the number actually observed (79).
The fraction of the number of stars in each age-class
was calculated using the main-sequence lifetimes
listed in \citet{scalo86}
and the mass-absolute magnitude-spectral type relations of \citet{ktg93}.

\subsection{Method}

\begin{figure}[t!]
\centerline{\psfig{figure=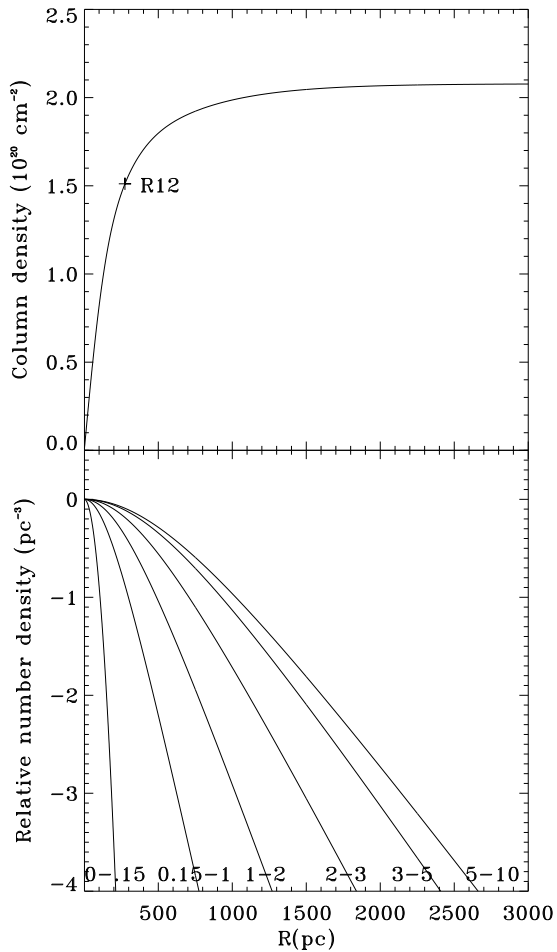,width=8.0cm}}
\caption{The relative number densities for each age group
as a function of {\it z}-height,
compared to the absorbing column through which they are observed.
{\it Top:} The total absorbing column through which
and object of {\it z}-height R is observed.
The $\tau=1$ point for band R12 is marked;
to achieve the same $\tau$ for band R47
requires a much greater column density.
{\it Bottom:} The relative number densities
for stars of a given age group at a {\it z}-height R.
Each curve has a mid-plane number density of unity.}
\label{fig:absorb}
\end{figure}

For each spectral-type/age-class bin,
the intrinisic luminosity distribution
was evaluated for $\Delta\log{L_X} = 0.1$ intervals.
For each distance interval, $\Delta r = 10$ pc,
from 0 to 3000 pc,
the value of $\log{L_X}$
was adjusted for $r^{-2}$ diminution,
absorption by the ISM,
and converted to count rates in a given band.
The probability of finding a star with flux $S$ in a given band,
as a function of S, was then interpolated to a
standard $\Delta\log{(S/\mbox{counts s}^{-1})}=0.1$
spacing,
and scaled by the size of the volume element
and the relative stellar density for that $z$-height.
The luminosity function for each spectral-type/age-class    
was then summed over $z$-height,
and scaled by the total number of stars

The vertical distribution of the ISM
was taken from \citet{lockman84}, 
and was assumed to be plane-parallel.
The absorption cross-sections were taken
from \citet{mm83}.
The absorption was calculated using the spectral energy distribution
described above.
Although the bulk of the stars lie beyond the bulk of the absorption
(see Figure~\ref{fig:absorb}),
only the R12 band reaches column densities for which $\tau=1$.
(The next most sensitive band is R17,
for which $\tau=1$ at $\sim5.5\times10^{20}$ cm$^{-2}$.)
Thus, the luminosity function is not too
sensitive to absorption by the ISM
at high Galactic latitudes.

The aggregate luminosity function for band R17
is shown in Figure~\ref{fig:lnls17}
for the north Galactic pole.
As might be expected,
the luminosity function is dominated by the older age classes because,
although their intrinsic luminosities may be an order of magnitude
smaller than the younger age classes,
they account for the bulk of the number of sources.
For similar reasons, the luminosity function
is dominated by the longer-lived spectral classes.
The high-count-rate end of the luminosity function
is strongly influenced by the younger age-classes
as they have larger intrinsic luminosities
and smaller scale-heights.

\begin{figure}[!p]
\centerline{\psfig{figure=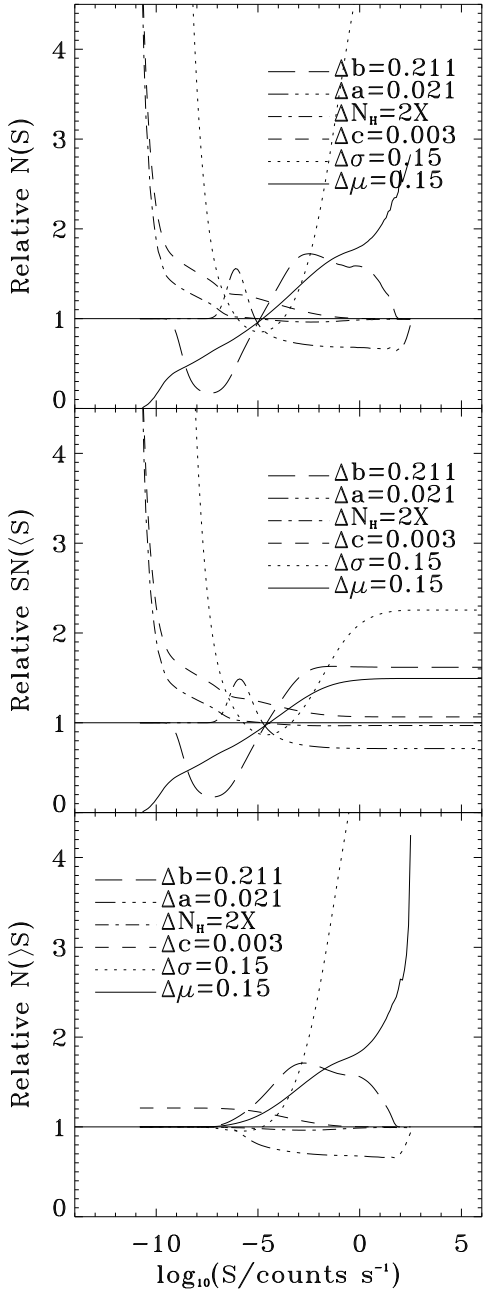,width=8.0cm}}
\caption{The effect of the formal uncertainties of the input parameters
on the luminosity function ({\it top}),
the unresolved flux ({\it middle}),
and the cumulative luminosity function ({\it bottom}),
shown as the ratio between the luminosity functions.
For the M(early), 1-2 Gyr age class
we show the R47 band luminosity function calculated with
a single parameter increased by $\Delta$,
where the $\Delta$ is the typical formal uncertainty,
divided by the fiducial luminosity function.}
\label{fig:uncer}
\end{figure}

Figure~\ref{fig:uncer} shows the effect
of the uncertainties of the parameters
on the luminosity function (top third),
unresolved flux (middle third),
and cumulative luminosity function (bottom third)
for a single spectral-type/age-class 
(early M, 1-2 Gyr) in band R47
(the most interesting band for cosmological purposes).
In this band, the effects of Galactic absorption are minor.
The parameter uncertainties are their typical formal uncertainties.
For typical survey limits,
$-4 < \log S(R47) < -2$,
the predicted unresolved flux
is uncertain to a factor of two due to the formal uncertainties.
The largest sources of uncertainty are the $\mu$ and $\sigma$
of the intrinsic luminosity distribution function.
Although the $b$ paramter of the $\log L_x$-$HR$ relation
also appears to contribute a significant uncertainty,
the true uncertainty is likely smaller than the value shown here,
which includes what may be a significant intrinsic scatter
around the relation.
Uncertainties in $c$ cause little change.

The true uncertainty in the amount of unresolved flux
is much less than that shown in Figure~\ref{fig:uncer};
as will be seen, the model can be checked against
the observed $N(S)$ over several orders of magnitude
in source flux,
and good agreement between the model and the observed $N(S)$
implies a smaller error in the unresolved flux.

\section{Comparisons}

The current model agrees fairly well with that of \citet{ghmr96}
except for the ``late M'' class.
For this class we find a factor of $\sim10$ more sources.
Only a small amount of this difference can be attributed to
the difference in luminosity distributions;
the luminosity distribution used here ($\mu=27.22$,$\sigma=0.49$)
is surprisingly similar to that derived from {\it Einstein} data
($\mu=27.7$, $\sigma\sim0.42$) by \citet{bmshr93}.
It is not clear whether the difference in number counts
for the late M class is due to a different IMF,
a different low-mass cut-off for X-ray emission
(0.08 M$_{\sun}$ was used here),
or merely a different mass-to-spectral-type
conversion.
All of these quantities are, of course, quite uncertain
for the low mass end of the main sequence.

\begin{deluxetable}{lccccccc}
\tablecolumns{8}
\tabletypesize{\footnotesize}
\tablecaption{Predictions
\label{tab:predict}}
\tablewidth{0pt}
\tablehead{
\colhead{Survey} &
\colhead{$(\ell,b)$} &
\colhead{Band} &
\colhead{Area} &
\colhead{ID} &
\colhead{Source} &
\colhead{Number} &
\colhead{Number} \\
\colhead{ } &
\colhead{2000} &
\colhead{ } &
\colhead{deg$^{2}$} &
\colhead{Completeness} &
\colhead{Limit} &
\colhead{Predicted} &
\colhead{Found} \\
\colhead{ } &
\colhead{ } &
\colhead{ } &
\colhead{ } &
\colhead{ } &
\colhead{$10^{-3}$ c/s} &
\colhead{ } &
\colhead{ } }
\startdata
\citet{bhcea96} & ( 96.4, 29.8) & 47 & 0.21 & 0.90 & .763 & 3.07 & 3 \nl
\citet{shg97}   & (149.5, 53.5) & 47 & 0.28 & 1.00 & .459 & 3.71 & 3 \nl
\citet{zmhea99} & (270.2,-51.8) & 47 & 0.20 & 0.84 & .309 & 3.73 & 4 \nl
\citet{mjmea98} & ( 85.0, 75.9) & 47 & 0.17 & 1.00 & .175 & 3.84 & 5 \nl
\citet{cwbea95} & (162.6, 51.1) & 47 & 2.20 & 0.57 & 1.42 & 11.11 & 9 \nl
\citet{bwe97}   & various       & 47 & 3.92 & 0.89 & 1.7  & 18.93 & 20 \nl 
\cutinhead{Comparison to measured log N-log S function}
\citet{kzaea99} & various       & 17 &      &      &   17 & 0.6\tablenotemark{a} & 0.64 \nl
                &               & 17 &      &      &  170 & 0.058 & 0.057 \nl
                &               & 17 &      &      & 1700 & 0.0032 & 0.0030 \nl
\enddata
\tablenotetext{a}{Value from the log N-log S function graphed in their paper.
We have assumed that this function is valid for $b=90\arcdeg$.}
\end{deluxetable}

\begin{figure}[bt!]
\centerline{\psfig{figure=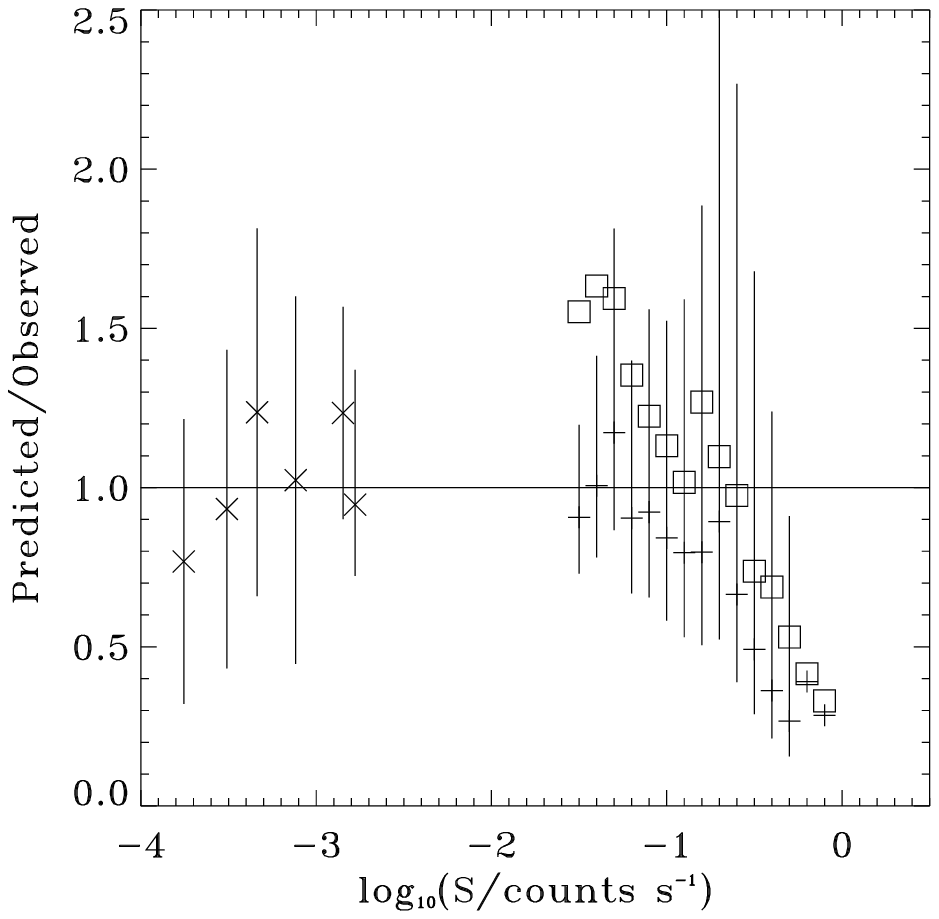,width=8.0cm}}
\caption{The predicted cumulative luminosity function
with measured values.
$\times$: values in band R47 from the deep surveys 
listed in Table~\ref{tab:predict},
+: (shown with error bars)
values in band R47 from the field IVac ($b=79\arcdeg$)
source list of \citet{atzk98}.
$\Box$: (not shown with error bars)
values in band R47 from fields I, II, III, and VI
of \citet{atzk98}.
These fields have $30\arcdeg<|b|<40\arcdeg$.
See text for further explanation.}
\label{fig:predict}
\end{figure}

We have checked our model against
the number of stellar point sources found in surveys
and against the stellar log N-log S relation of \citet{kzaea99}.
These comparisons are listed in Table~\ref{tab:predict}
and shown in Figure~\ref{fig:predict}.
The predicted number listed in the Table
is the number of sources with flux
above the survey limit at the $(\ell,b)$
position of the survey, in degrees$^{-2}$,
multiplied by the survey area
and the fraction of sources for which
identifications have been made
(the ``ID completeness'').

As can be seen,
the model predicts well the band R47 source counts
for fluxes below $\log_{10}S\sim-2.5.$
At higher fluxes we can compare our model to the wide-field
(each field has $\sim144$ square degrees) data of \citet{kzaea99} 
which is complete in band R17 to 0.03 counts s$^{-1}$,
(and thus must be complete to at least that level in the sub-band R47).
We find that in band R47, that for their field IVac ($b=79\arcdeg$),
the model is consistent with observations for
fluxes below 0.1 counts s$^{-1}$.
However, below that threshold,
the model over-predicts at low latitudes
($30\arcdeg<|b|<40\arcdeg$, their fields I, II, III, and VI).
Above this threshold, the model under-predicts
the source counts at all latitudes.
A similar trend is seen in the band R17 source counts;
the model somewhat under-predicts the counts at
high Galactic latitudes and over-predicts
at low Galactic latitudes for fluxes greater than 0.03 counts s$^{-1}$.
Given the shape of the luminosity functions (Figure~\ref{fig:lnls17}),
this error is likely due to the poorly determined
$\mu$ and $\sigma$ of the youngest age group.
({\it Nota Bene:} the caveat given in \S2
concerning the Hyades luminosity function,
as well as our use of a $\log L_x$-$HR$ relation
for the M stars that may be too hard,
decreasing the effect of absorption, \S2.1.)

The problems with the model at bright fluxes
do not cast serious doubt upon our prediction
at lower fluxes;
as can be seen from Figure~\ref{fig:uncer},
the brighter fluxes are much more sensitive
to uncertainties in $\mu$ and $\sigma$,
and the brighter fluxes are much more sensitive
to the younger age groups.
It should also be noted that the uncertainties
in the unresolved flux at the brighter fluxes
will be significantly smaller than those of the source counts.

\section{Model Description}

It is instructive to see how the luminosity function
changes with detection band,
Galactic latitude,
and Galactic longitude.

\begin{figure}[tb!]
\centerline{\psfig{figure=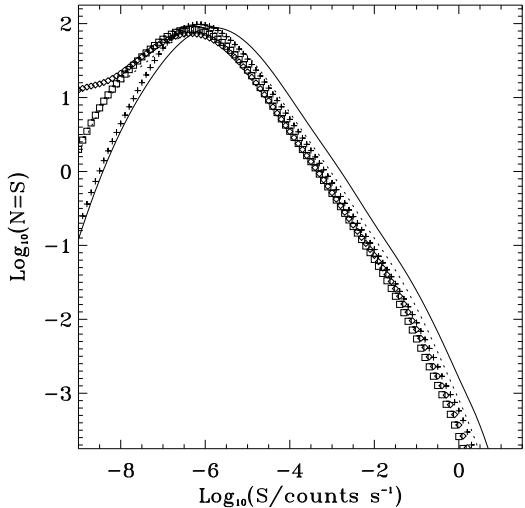,width=8.0cm}}
\caption{The luminosity function for different bands
in the direction of the north Galactic pole.
{\it Solid line:} band R17,
+: band R12,
$\Box$: band R45,
$\Diamond$: band R67,
{\it Dotted line:} band R47 (almost indistinguishable from band R45).}
\label{fig:lnls_bands}
\end{figure}

Figure~\ref{fig:lnls_bands} shows the stellar luminosity function
at the north Galactic pole for several \rosat\ bands.
The R12 band is substantially different from the R45 band
due both to the different instrumental response
(which produces a horizontal shift in the curves)
and the differing effects of absorption
(which increases the number counts below the peak
and decreases them above the peak,
though not at the highest fluxes).
The anomolous behavior of band R67
is due to the discontinuity of the $\log L_x$-$HR$
relation at $HR=-0.86$;
the rate of decrease of the band R67 flux with decreasing $HR$
is greater than for the other bands,
thus the discontinuity is more pronounced for band R67
than for the other bands.
As the anomolous behavior occurs two orders of magnitude
below the peak of the luminosity function,
it does not cause significant errors.

\begin{figure*}[t!]
\centerline{\psfig{figure=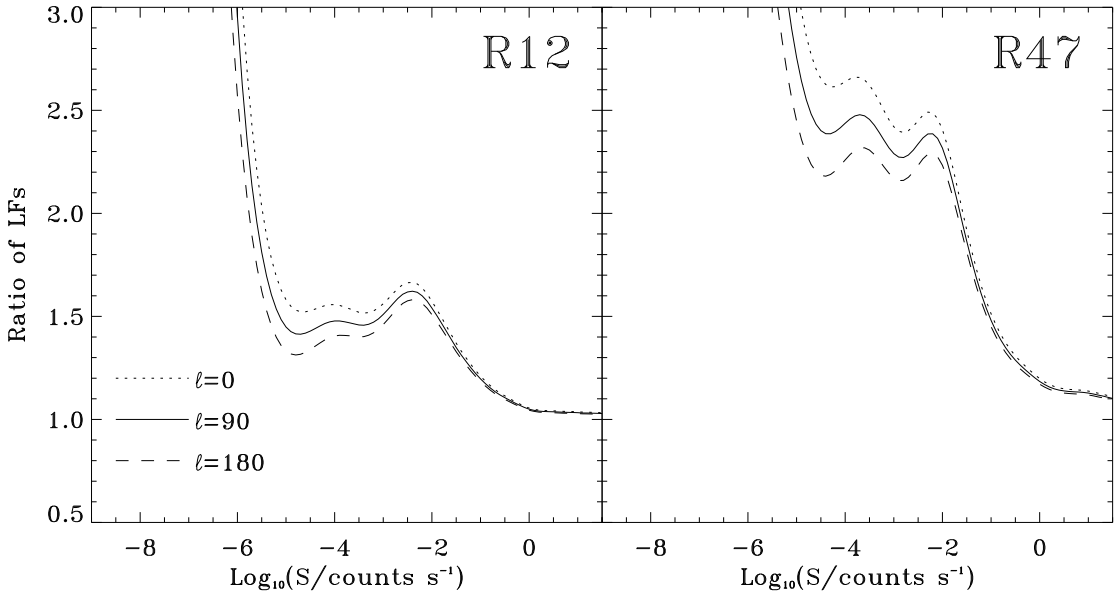,width=16.0cm}}
\caption{The ratio of the luminosity function at $b=30\arcdeg$
to the luminosity function at $b=90\arcdeg$.
{\it Solid:} The ratio for $\ell=90\arcdeg$,
{\it Dotted:} the ratio for $\ell=0\arcdeg$,
{\it Dashed:} the ratio for $\ell=180\arcdeg$.}
\label{fig:lats}
\end{figure*}

Figure~\ref{fig:lats} shows the ratio of the stellar luminosity function
at $b=30\arcdeg$ to that at the Galactic pole for the R12 and R47 bands.
We have assumed a plane-parallel absorption model.
For typical deep \rosat\ surveys,
there is a factor of $\sim2.5$ greater number of sources
at $b=30\arcdeg$ than at $b=90\arcdeg$ in band R47.
The increase in number counts for band R12 is less severe
due to absorption.
As can be seen from the Figure,
at $b=30\arcdeg$ 
the effect of the Galactic radial density gradient is negligible.

\section{Results}

\begin{figure}[tb!]
\centerline{\psfig{figure=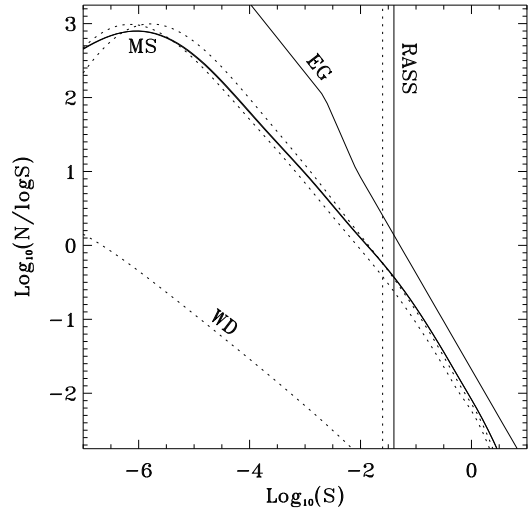,width=8.0cm}}
\caption{The luminosity function for main-sequence stars,
white dwarfs, and extragalactic objects,
for band R12 (dotted) and band R47 (solid).
The relations are shown for a total $N_H = 5\times10^{19}$ cm$^{-2}$
(further to the right)
and for $N_H = 2.1\times10^{20}$
(further to the left).
For band R47, the relations for the different values of $N_H$
are indistinguishable.
The band R47 relation for white dwarfs is not shown,
but would be significantly below the R12 relation.
The point source detection limit for the \rosat\ All-Sky Survey
is also shown.}
\label{fig:finalp}
\end{figure}

Figure~\ref{fig:finalp} shows the luminosity functions
in PSPC bands R12 (dotted) and R47 (solid)
for main-sequence stars, white dwarfs\footnote{
The white dwarf luminosity function for single white dwarfs
is shown to demonstrate its inconsequentiality.
It was calculated using the data of \citet{fspbg96}
with the assumption that the intrinsic X-ray luminosity
of white dwarfs extends down to at least $\log{L_X}=27.5$},
and the extragalactic luminosity function
derived by \citet{hea98}.
For band R47, above the RASS detection limits,
stellar sources account for one third to one half of point sources,
in agreement with studies
such as those listed in Table~\ref{tab:predict},
while for band R12,
the fraction depends upon the total absorbing column.

\begin{deluxetable}{lccccc}
\tablecolumns{6}
\tabletypesize{\footnotesize}
\tablecaption{Unresolved Stellar Flux
\label{tab:unres_flux}}
\tablewidth{0pt}
\tablehead{
\colhead{{\it ROSAT}} &
\colhead{Energy} &
\colhead{{\it RASS}} &
\multicolumn{3}{c}{Unresolved Flux} \\
\colhead{Band} &
\colhead{Range\tablenotemark{a}} &
\colhead{PSDL\tablenotemark{b}} &
\colhead{} &
\colhead{} &
\colhead{} \\
\cline{4-6} \\
\colhead{} &
\colhead{keV} &
\colhead{counts s$^{-1}$} &
\colhead{10$^{-6}$ counts s$^{-1}$ arcmin$^{-2}$} &
\colhead{10$^{-14}$ erg cm$^{-2}$ s$^{-1}$ deg$^{-2}$} &
\colhead{keV cm$^{-2}$ s$^{-1}$ sr$^{-1}$ keV$^{-1}$} }
\startdata
R12 & 0.11-0.284 & 0.025 & 6.85 & 4.66 & 0.55 \nl
R45 & 0.44-1.21  & 0.02  & 4.76 & 31.3 & 0.83 \nl
R67 & 0.73-2.01  & 0.02  & 4.91 & 26.9 & 0.42 \nl
\enddata
\tablenotetext{a}{Energy at 10\% of the peak response.}
\tablenotetext{b}{Point Source Detection Limit.}
\end{deluxetable}

\begin{figure}[tb!]
\centerline{\psfig{figure=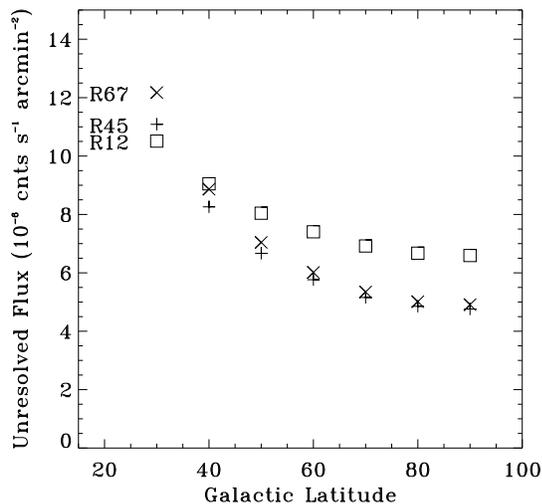,width=8.0cm}}
\caption{The contribution of unresolved stellar sources
to the \rass\ as a function of Galactic latitude.
It is assumed that the point source removal limits were
0.025 counts s$^{-1}$ in band R12
and 0.02 counts s$^{-1}$ in band R45 and band R67.}
\label{fig:unres}
\end{figure}

Figure~\ref{fig:unres} shows the contribution to the \rass\
of unresolved stellar sources as a function of Galactic latitude.
Table~\ref{tab:unres_flux} shows the model prediction
for the contribution of the unresolved Galactic stars
to the diffuse X-ray background
at the north Galactic pole for $N_H=2.1\times10^{20}$ cm$^{-2}$.
The conversion of count rates to energy
use the spectral shape of the unresolved flux
and the energy ranges listed in the table.
At the \rass\ point source detection limit in band R45
the unresolved flux has R12:R45:R67 =1.51:1:1.09,
and those ratios are similar for the unresolved flux
at the \rass\ point source detection limits in the other bands.
The unresolved flux in the R47 band (0.44-2.04 keV)
is $3.64(3.51)\times10^{-13}$ erg cm$^{-2}$ s$^{-1}$ deg$^{-2}$
at the \rass\ band R45(R67) point source detection limit.

In bands R12, R45, and R67,
the unresolved stellar sources can account for, 
at most, 2\%, 10\% and 51\% of the flux 
that is thought not to be due to the Local Hot Bubble
or the unresolved extragalactic point sources
(essentially the component that \citet{ks2000}
attribute to the Galactic halo,
but which can be distributed between the Galactic halo
and diffuse extragalactic emission).
The variation of the contribution from unresolved
stellar sources with Galactic latitude
would be unnoticeable in the \rass\
given the uncertainty in the \rass\ and,
more importantly, the variable effects of Galactic absorption.

The deep band R47 surveys 
discussed in \S3.2 suggest that
the predicted unresolved flux
for surveys at high Galactic latitudes
with limits $\log S(R47) < -2.5$ is correct.
At higher fluxes the comparison of the model 
with observations is more equivocal,
but the high Galactic latitude source counts
in band R47 do seem to be correct,
(at least for $S\lesssim0.3$ counts s$^{-1}$)
and thus the unresolved flux should be correct.
Due to the uncertainties introduced by the younger stars,
the model is not adequate at lower Galactic latitudes ($|b|<30\arcdeg$),
particularly for the high flux source counts
as these are dominated by younger stars.
It should also be noted that both the younger stars
and the Galactic absorption
will be distributed in a much more clumpy manner
than used by this model,
so a great deal of field to field
variation is to be expected.

\acknowledgements

We would like to thank T. Fleming for several useful 
discussions about stellar X-ray emission,
and for his patience as a referee,
J.H.M.M. Schmitt for a discussion of the $\log L_x$-$HR$ relations,
P. Guillout for guidance on the stellar distribution,
and S. Drake for discussions about selection effects
in the data from which we constructed
our empirical model of the stellar X-ray emission.

\bibliographystyle{apj}
\bibliography{apjmnemonic,biblio}
\end{document}